\documentclass[conference]{IEEEtran}
\def\IEEEtitletopspace{18pt}
\IEEEoverridecommandlockouts

\usepackage{cite}
\usepackage{amsmath,amssymb,amsfonts}
\usepackage{algorithmic}
\usepackage{graphicx}
\usepackage{textcomp}
\usepackage{xcolor}
\usepackage{booktabs}
\usepackage{mathtools}
\usepackage{multirow}
\usepackage{url}
\usepackage[hidelinks]{hyperref}

\makeatletter
\let\MYcaption\@makecaption
\makeatother

\usepackage[font=footnotesize]{subcaption}

\makeatletter
\let\@makecaption\MYcaption
\makeatother

\def\BibTeX{{\rm B\kern-.05em{\sc i\kern-.025em b}\kern-.08em
    T\kern-.1667em\lower.7ex\hbox{E}\kern-.125emX}}

\DeclareMathOperator*{\argmin}{arg\,min}

\begin{document}

\title{
Safe and Efficient CAV Lane Changing using Decentralised Safety Shields
}

\author{
\IEEEauthorblockN{
Bharathkumar Hegde \qquad
M\'elanie Bouroche
}
\IEEEauthorblockA{
School of Computer Science and Statistics, Trinity College Dublin, Ireland \\
\{hegdeb, Melanie.Bouroche\}@tcd.ie
}
}

\maketitle
\begin{abstract}

Lane changing is a complex decision-making problem for Connected and Autonomous Vehicles (CAVs) as it requires balancing traffic efficiency with safety. Although traffic efficiency can be improved by using vehicular communication for training lane change controllers using Multi-Agent Reinforcement Learning (MARL), ensuring safety is difficult.
To address this issue, we propose a decentralised Hybrid Safety Shield (HSS) that combines optimisation and a rule-based approach to guarantee safety. Our method applies control barrier functions to constrain longitudinal and lateral control inputs of a CAV to ensure safe manoeuvres. Additionally, we present an architecture to integrate HSS with MARL, called MARL-HSS, to improve traffic efficiency while ensuring safety. 
We evaluate MARL-HSS using a gym-like environment that simulates an on-ramp merging scenario with two levels of traffic densities, such as light and moderate densities. The results show that HSS provides a safety guarantee by strictly enforcing a dynamic safety constraint defined on a time headway, even in moderate traffic density that offers challenging lane change scenarios. Moreover, the proposed method learns stable policies compared to the baseline, a state-of-the-art MARL lane change controller without a safety shield. Further policy evaluation shows that our method achieves a balance between safety and traffic efficiency with zero crashes and comparable average speeds in light and moderate traffic densities.
\end{abstract}


\section{Introduction}

Autonomous Vehicles (AVs) were expected to be commercially available by 2020, but recent reports suggest that wider adoption of AVs can only be expected after 2030 or beyond due to societal, regulatory, and technical challenges \cite{agrawal_building_2023}. Complex technical problems, such as localisation, mapping, perception, route planning, and motion control, are yet to be solved to enable commercial AV deployments \cite{ma_artificial_2020}.
Among them, safe and efficient lane change in dynamic traffic conditions is one of the difficult problems because AV controllers must analyse the environment and actuate a control decision in a very short time period, almost instantaneously. Moreover, improper lane change can lead to a collision that could cause significant damage and loss of lives. Safe and strategic lane changes can improve traffic efficiency in scenarios that require mandatory lane changes, such as highway merging \cite{marinescu_-ramp_2012}. 

Using vehicular communication (V2X), Connected Autonomous Vehicles (CAVs) can collaborate to analyse the environment and make efficient control decisions. In particular, CAVs can make efficient lane change decisions using Multi-Agent Reinforcement Learning (MARL)  \cite{hegde_design_2022}, \cite{liu_systematic_2023}. Many forms of MARL have been applied for designing lane change controllers including Deep Q-Networks (DQNs) \cite{yu_distributed_2020}, \cite{dong_space-weighted_2021}, \cite{chen_graph_2021}, and Actor-Critic Networks (ACN) \cite{zhou_multi-agent_2022}, \cite{hou_decentralized_2021}, \cite{chen_deep_2023}. Among them, MARL-CAV is an open-source state-of-the-art MARL lane change controller designed for CAVs \cite{chen_deep_2023}. MARL-CAV formulates a Partially Observable Markov Decision Process (POMDP) to train lane change controllers in a highway merging section. This approach uses a predication-based priority assignment to avoid collisions, but the safety of the lane change is not guaranteed. Moreover, this priority-based safety layer requires a centralised server to monitor and communicate with CAVs, which hinders its scalability. 

In our previous work, we found that Control Barrier Functions (CBFs) are suitable to ensure the safety of CAV lane change controllers \cite{hegde_multi-agent_2024}.
CBFs have also been applied to ensure the safe operation of single-agent RL-based AV controllers \cite{cheng_end--end_2019}. This approach demonstrates the longitudinal safety of a vehicle using simple kinematic equations. Furthermore, Han et al. demonstrate that CBFs can be used to define longitudinal and lateral safety constraints for MARL-based CAV controllers \cite{han_multi-agent_2023}. Their CBF formulation, however, can be computationally expensive as reference trajectories are calculated for all possible behavioural control actions.  
To improve computational efficiency while constraining both longitudinal and lateral control of an AV,  the use of Higher-Order CBFs (HOCBFs) has been proposed \cite{wang_ensuring_2022}. Notably, the lateral constraints in this formulation assume that the rate of heading change is directly controlled. While this assumption is appropriate for simple robotic systems or simulation setups, AVs are typically controlled by adjusting the steering angle rather than the heading directly. Moreover, existing CBF formulations define safety constraints to maintain a constant distance from obstacles. However, using a dynamic safe distance based on time headway can improve traffic efficiency \cite{garg_can_2023}.

Overall, the existing work does not fully address the challenges of making safe and efficient lane changes.
While MARL lane change controllers can be trained to perform efficient lane changes, they do not guarantee safety. 
Although CBFs can be integrated with MARL to ensure safety, existing approaches do not fully address safe lateral control and do not support dynamic safety distances. In our previous work \cite{hegde_safe_2024}, we proposed a theoretical framework, CBF-CAV, that integrates CBF constraints with MARL to ensure safe execution of the behavioural lane change decisions, but it was not empirically evaluated. This paper significantly extends the CBF-CAV framework with the following  contributions: 
\begin{itemize}
    \item We propose a decentralised Hybrid Safety Shield (HSS) to constrain CAV controls using the CBFs while respecting their control limits. This shield is designed to effectively ensure the safety of both longitudinal and lateral control inputs based on a dynamic safety constraint. 
    
    \item We define a MARL-HSS architecture to integrate MARL with the HSS for efficiently training CAV lane change controllers with a safety guarantee from HSS. With MARL-HSS, CAV lane change controllers can be trained to make efficient decisions while ensuring safety.
    
    \item The MARL-HSS is comprehensively evaluated in a simulation of a highway on-ramp scenario with different traffic densities. The results show that MARL-HSS outperforms the state-of-the-art MARL lane change controller by effectively balancing between safety and efficiency.
\end{itemize}

The proposed method, MARL-HSS, considers pure CAV traffic, with vehicles capable of sharing their state without any communication delays or packet loss. This assumption avoids the risk of unknown behaviour from other vehicles, which is not handled in the current safety shield.

\section{Background}
This section introduces the background knowledge on AV control hierarchies, MARL, vehicle model, and CBFs to provide a context for the proposed method.

\subsection{Control hierarchies}
The control decisions of AVs can be separated into four hierarchical control levels such as route planning, behavioural, motion planning, and vehicular control \cite{paden_survey_2016}. 
The route planning layer first identifies a feasible route to the destination provided by the user using road network information.
To move along the defined route, a behavioural layer makes discrete decisions to change lanes or follow the current lane. The motion planning layer then identifies the control references to execute the desired driving manoeuvre to implement each behavioural decision. These driving manoeuvres are then actuated by the vehicular controller. 
Among them, control decisions from high-level behavioural and low-level motion planning layers influence the safety and efficiency of lane changes. Therefore, this article mainly focuses on designing these layers in the AV control hierarchy. Moreover, since connectivity among AVs enables collaborative control decisions to improve traffic efficiency, this article considers CAVs for their potential benefits \cite{matin_impacts_2022}. 

\subsection{MARL for on-ramp merging} 
\label{subsec:marl}
CAVs can use MARL to make efficient lane change decisions at a behavioural level based on its state and the states of surrounding vehicles to improve the traffic flow. MARL-CAV is a state-of-the-art decentralised MARL formulation for on-ramp merging, where each vehicle observes only a part of the environment \cite{chen_deep_2023}. Therefore, a POMDP can be formulated for a multi-agent system of $\mathcal{N}$ vehicles described by a tuple $( \left\{ \mathcal{S}_i, \mathcal{A}_i, \mathcal{R}_i \right\}_{i\subseteq \mathcal{N}}, \mathcal{T})$.   

The \emph{state space}, $\mathcal{S}_i$, of a vehicle $i$ consists of following state variables,
\begin{itemize}
    \item \verb+is_present+~: a binary variable representing the presence of a vehicle within the considered scenario. 
    \item $x$~: The longitudinal position of the vehicle.
    \item $y$~: The lateral position of the vehicle.
    \item $v_x$~: The longitudinal velocity of the vehicle.
    \item $v_y$~: The lateral velocity of the vehicle.
    \item $\psi$~: The vehicle heading with respect to the road.
\end{itemize}

The state space also includes the states of $n$ surrounding vehicles within the perception range of a CAV. While the state variables of the ego vehicle are captured with respect to a global coordinate system, the observed vehicle's state variables are relative to the ego vehicle. A CAV can measure its own state from onboard sensors such as LIDAR, RADAR, Camera, GPS, and IMU. These sensors are assumed to provide accurate state estimation. The V2X communication interface can be used to capture the states of the surrounding vehicles.
The overall multi-agent state space is defined as a Cartesian product of the individual states $\mathcal{S} = \mathcal{S}_0 \times \mathcal{S}_1 \times \mathcal{S}_2 \times~...~\times \mathcal{S}_{\mathcal{N}}$.

The \emph{action space}, $\mathcal{A}_i$, consists of five discrete variables representing a specific lane change behaviour such as \{\verb+right+, \verb+left+, \verb+follow lane+, \verb+speed up+, \verb+slow down+\}. The behavioural layer chooses one of these high-level actions. The overall multi-agent action space is defined as joint actions from the individual AVs, $\mathcal{A} = \mathcal{A}_0 \times \mathcal{A}_1 \times \mathcal{A}_2 \times~...~\times \mathcal{A}_{\mathcal{N}}$.

The \emph{reward} function, $\mathcal{R}_i$, rewards an individual CAV for achieving individual and collaborative goals. The reward for a CAV $i$ can be defined as
\begin{equation}
    r_{i} = w_\text{c} r_\text{c} + w_\text{s} r_\text{s} + w_\text{h} r_\text{h} + w_\text{m} r_\text{m}
    \label{eq:reward}
\end{equation}
where a CAV is rewarded for avoiding collision with $r_\text{c}$, maintaining desirable speed with $r_\text{s}$, maintaining desirable headway with $r_\text{h}$, and minimising the merging wait time with $r_\text{m}$. Each of these rewards is multiplied with a weight $w_*$ that can be tuned to prioritise the different objectives. 
The individual rewards from the $n$ observed vehicles within the perception range are combined to consider collaborative goals such as traffic efficiency. Therefore, the reward function, $\mathcal{R}_i$, is defined as
\begin{equation}
    R_{i} = \frac{1}{n}\sum_{j=0}^{n} r_{j}
\end{equation}

The transition probability, $\mathcal{T}$, characterises the dynamics of the CAV multi-agent system. To navigate CAVs based on control inputs, a vehicle model (explained in section~\ref{subsec:vehicle_model}) is used. 
This MARL formulation is integrated with a safety shield proposed in Section~\ref{sec:proposed_approach}.

\subsection{Vehicle model}
\label{subsec:vehicle_model}
It is difficult to model AV dynamics as they are affected by a variety of factors ranging from vehicle control to weather conditions. For simplicity, the kinematic bicycle model is often used to model AV motion \cite{polack_kinematic_2017}, and we demonstrate the proposed safety shield using this model. 

\begin{figure}[tbp]
    \centering
    \includegraphics[width=\linewidth]{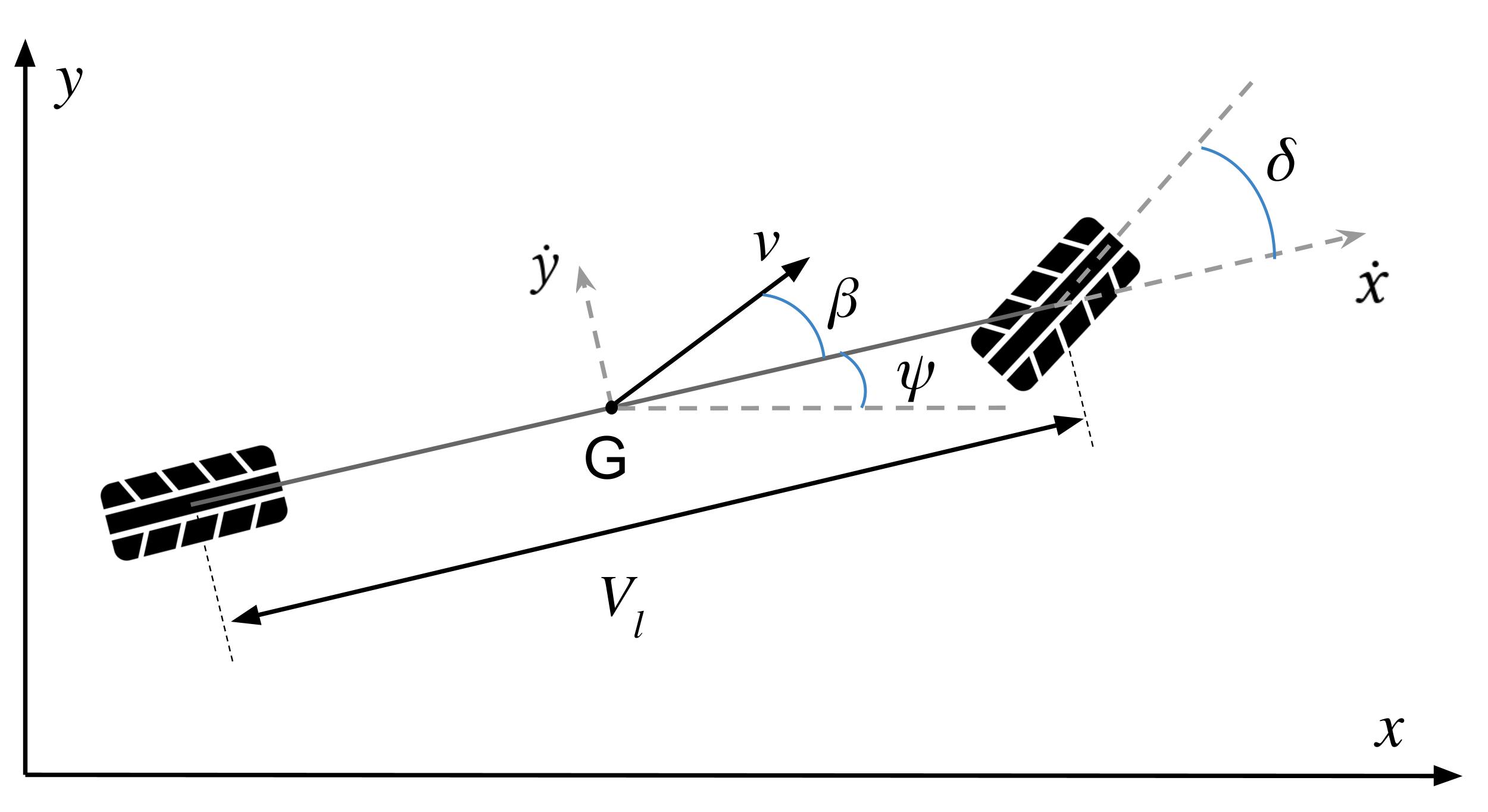}
    \caption{Kinematic bicycle model}
    \label{fig:kinematic_model}
    \vspace{-0.25cm}
\end{figure}

The kinematic bicycle model considers the two front wheels as one wheel and the same for the back wheels, as illustrated in Fig.~\ref{fig:kinematic_model}. The following equations define the vehicle model based on the state variables defined previously (\ref{subsec:marl}),
\begin{align}
    \label{eq:kinematic_model}
    \begin{split}    
    \dot{x} = v_x,~~~~
    \dot{y} = v_y,~~~~
    \dot{v}_x = a\cos{(\psi + \beta)},\\
    \dot{v}_y = a\sin{(\psi + \beta)},~~~~
    \dot{\psi} = \frac{2~v}{V_l}\sin{\beta},~~~~
    \end{split}
\end{align}
where $\beta$ is a slip angle at the centre of gravity evaluated based on the steering angle input $\delta$:
\begin{align}
    \label{eq:slip_angle}
    \beta &= \tan^{-1} \left( \frac{1}{2}\tan\delta \right)
\end{align}
This vehicle model is used to integrate MARL with the CBF constraints formulated within the proposed HSS in Section~\ref{subsec:marl_hss}. 

\subsection{Control Barrier Functions for motion planning}

A CBF can be used to constrain a non-linear control system such as the motion planning layer in a vehicle control to ensure safety \cite{ames_control_2019}.
Consider a discrete time non-linear control system defined by the following transition dynamics
\begin{equation}
    \label{eq:nonlin_control}
    s_{t+1} = f(s_t) + g(s_t)u_t
\end{equation}
where the next state $s_{t+1}$ is defined based on unactuated dynamics $f:S\rightarrow S$, and actuated dynamics $g: S \rightarrow \mathbb{R}^{n,m}$, $n$ and $m$ are the number of variables in the state space $S$ and action space $U$ respectively, $s_t \in S$ is the system state, and $u_t \in U$ is the control action at time step $t$. 
The $f$ and $g$ are defined based on known system dynamics and they are locally Lipschitz continuous, in other words, continuous functions limited by a maximum rate of change. 

Consider a safe set $C$ defined as the super-level set of the continuously differentiable function $h: S \rightarrow \mathbb{R}$ 
\begin{equation}
\label{eq:safe_set}
    C : \{s_t \in S: h(s_t) \geq 0\}
\end{equation}

To ensure the safety of the control system (\ref{eq:nonlin_control}), the safe set $C$ must be forward invariant, which means that safe actions can be defined for each state $s_t \in C$ such that the system continues to stay in $C$. 
The safe set $C$ is forward invariant if the function $h$ is a \emph{Control Barrier Function} (CBF) such that there exists $\eta \in [0,1]$ for all $s_t \in C$ satisfying the following equation
\begin{equation}
    \label{eq:cbf}
    \sup_{u_t \in U} \left[h\left(f(s_t) + g(s_t)u_t\right) + (\eta - 1)h(s_t)\right] \geq 0
\end{equation}
where $\eta$ is the coefficient of zero order control that defines the magnitude at which the system is pushed within the safe set $C$ \cite{cheng_end--end_2019}.
Using smaller values of $\eta$ can enforce the constraints strictly, whereas higher values can relax the constraints. Therefore, $\eta$ represents how strongly the barrier function pushes the states inwards within $C$.
The existence of a CBF implies that for all $s_t \in C$, there exist $u_t$ such that $C$ is forward invariant \cite{ames_control_2019}. Therefore, the goal is to find a minimal safe action $u_t$ that satisfies (\ref{eq:cbf}) to ensure the safety of a control system (\ref{eq:nonlin_control}).

To define a CBF $h$, let us consider the affine barrier function of the form
\begin{equation}
    \label{eq:affine_fn}
    h(s_t) = p^{\text{T}}s_t+q
\end{equation}
where $p \in \mathbb{R}^n$ and $q \in \mathbb{R}$ are the parameters used to define a safety constraint $h$ on the state $s_t$. Combining the affine barrier function with the condition (\ref{eq:cbf}), the following constraint can be defined for the control action $u_t$,
\begin{equation}
    \label{eq:cbf_invariance}
    -p^{\text{T}}g(s_t)u_t \leq p^{\text{T}}f(s_t) + p^{\text{T}}(\eta - 1) s_t + \eta q
\end{equation}

To consider multiple safety constraints defined using CBFs, $C$ can be considered as the intersecting half spaces defined by $k$ affine barrier functions \cite{wang_ensuring_2022}. The affine constraint on $u_t$ can be defined by stacking all the constraints.
\begin{align}
    \label{eq:opt_constraint}
    \begin{split}
        Au_t &\leq b,\\
        \text{where, ~~} A = [-p_1^{\text{T}}g(s_t),...&,-p_k^{\text{T}}g(s_t)] \\
        b = [b_1,..., b_k] \text{, with} ~ b_i =&~ p_i^{\text{T}}f(s_t)~ + \\
        &p_i^{\text{T}}(\eta - 1) s_t + \eta q_i
    \end{split}
\end{align}

This constraint can be used to reformulate the CBF given by Eq.~(\ref{eq:cbf}) into the following optimisation problem with a slack variable $\epsilon$ in the safety condition and a large constant, $K_{\epsilon}$ that penalizes safety violations.
\begin{align}
\label{eq:opt_qp}
\begin{split}
    u_t = & \argmin_{u_t} ||u_t||_2 + K_{\epsilon}\epsilon\\
          & \text{s.t}~~ Au_t \leq b,
\end{split}
\end{align}
This optimisation problem can be efficiently solved in each time step using Quadratic Program (QP) \cite{boyd_convex_2004}.

If the CBF $h$ is defined for state variables such as $x$ and $y$, control input $a_t$ cannot be effectively constrained using the above optimisation problem (\ref{eq:opt_qp}) because the constraints are effective only when $A$ is non-zero. The parameter $A$, defined in Eq.~(\ref{eq:opt_constraint}), consists of $-p^{\text{T}}_i g(s_t)$, where $g(s_t)$ can be zero if a state variable is not affected by the control input $u_t$ directly. Since acceleration control $a_t$ changes the velocity $v$ of the vehicle in the immediate next step, $g(s_t)$ will be a non-zero value for the CBF constraints defined for velocities $v_x$ and $v_y$.  However, $g(s_t)$ will be zero for the constraints defined on the position of a vehicle, such as $x$ and $y$. Since the proposed Hybrid Safety Shield (\ref{sec:proposed_approach}) defines constraints on the vehicle's position, it uses the velocity $v$ as a control input.

\section{Proposed approach}
\label{sec:proposed_approach}
In this section, a Hybrid Safety Shield formulation for a multi-agent system of CAVs is defined using CBF constraints. First, the HSS formulation is presented to constrain control inputs of the motion planning layer to ensure safety. Then, MARL-HSS architecture is presented to integrate HSS with a MARL behavioural lane change controller. The MARL-HSS aims to maximise traffic efficiency while ensuring safety.   

\subsection{Hybrid Safety Shield}
\label{subsec:hss}
We propose a Hybrid Safety Shield that constrains longitudinal and lateral control inputs while considering their physical limits using CBFs to ensure safety. In the HSS, an optimisation problem derived from Eq.~(\ref{eq:opt_qp}) is solved to ensure safe longitudinal motion and a rule-based method is used to ensure safe lateral motion. The optimisation approach uses CBF constraints defined based on the longitudinal time headway with respect to a leading vehicle. The rule-based approach depends on longitudinal time headways with respect to the vehicles in the target lane. Moreover, the control input's physical limits are integrated into the optimisation problem.

\begin{figure}[tbp]
\centering
    \includegraphics[width=\linewidth]{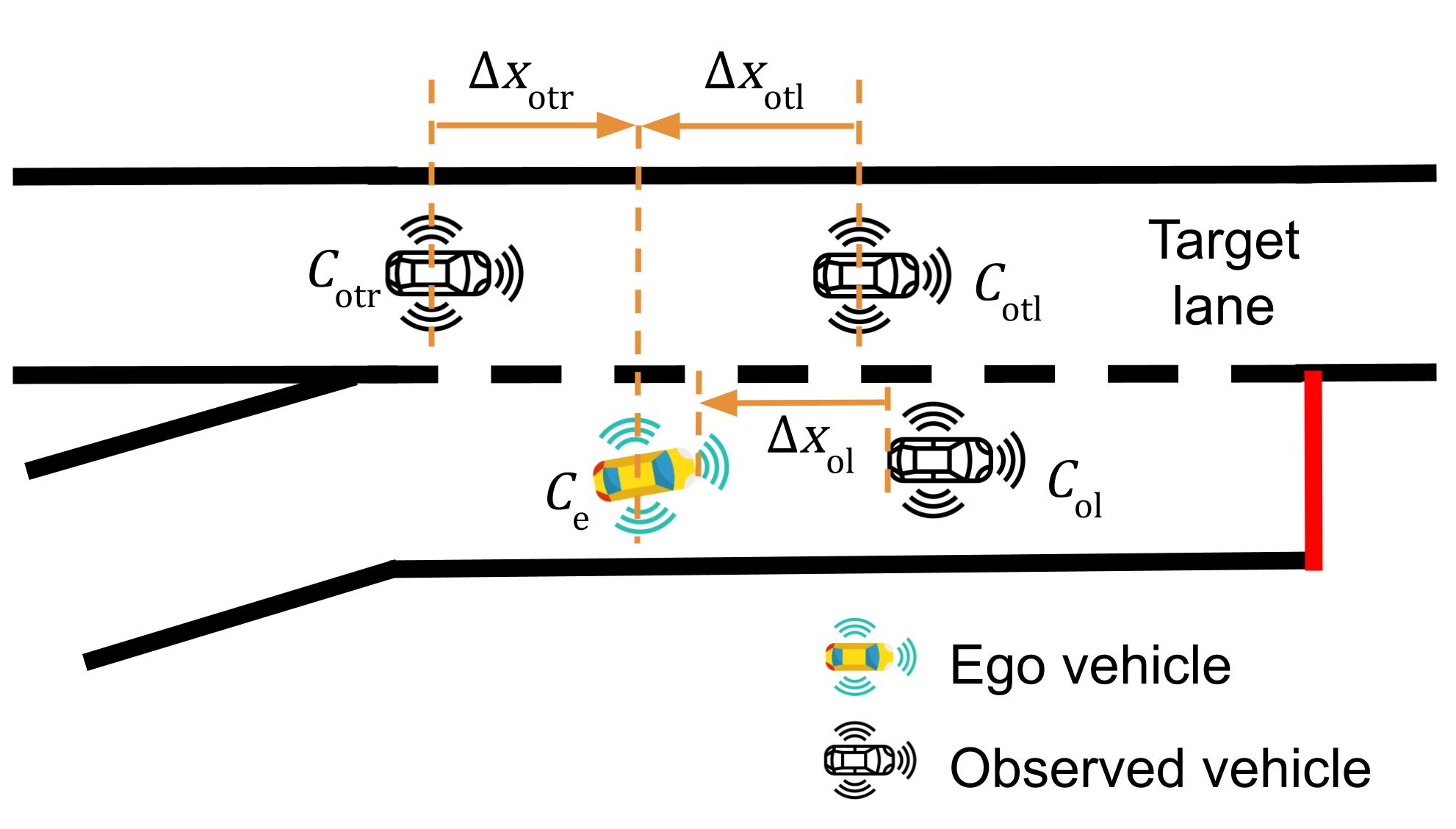}
    \caption{Topology to identify surrounding vehicles}
    \label{fig:topology_a}
    \vspace{-0.25cm}
\end{figure}

Identifying the surrounding vehicles is important to make safe manoeuvres. Therefore, we define a topology to identify the surrounding vehicles in a scenario with two lanes. This topology identifies three observed CAVs $C_\text{o*}$, namely the \emph{leading} vehicle $C_\text{ol}$, the \emph{target leading} vehicle $C_\text{otl}$, and the \emph{target rear} vehicle $C_\text{otr}$. The vehicle immediately in front of the ego vehicle in the same lane is considered to be the leading vehicle. While changing lanes, a vehicle moving immediately in front of the ego vehicle in the target lane is considered the target leading vehicle. A rear vehicle in the target lane is considered to be the target rear vehicle. In a merging section with two lanes, the topology of the surrounding vehicles is shown in Fig.~\ref{fig:topology_a}. The longitudinal distances from the surrounding vehicles $C_\text{ol}$, $C_\text{otl}$, and $C_\text{otr}$ are defined as $\Delta x_\text{ol}$, $\Delta x_\text{otl}$, and $\Delta x_\text{otr}$ respectively. The surrounding vehicle states are used in the safety constraints defined for longitudinal and lateral vehicle controls. 

\subsubsection{Longitudinal control}

Let us define a control system  (\ref{eq:nonlin_control}) for the longitudinal motion of a vehicle as follows
\begin{equation}
    \label{eq:nonlin_control_lonav}
    x_{t+1} = x_t + v \cos{(\psi + \beta)}
\end{equation}
where $x_t$ is the position along the longitudinal direction at time step $t$, and  $\psi$ and $\beta$ are the heading angle and slip angle defined in the vehicle model (\ref{eq:kinematic_model}). This Eq.~(\ref{eq:nonlin_control_lonav}) can be extended to a multi-agent system of CAVs in a matrix form as follows,

\begin{align}
    \label{eq:nonlin_control_loncav_matrix}
    \begin{bmatrix}
    x_{\text{e}}\\
    x_\text{o}
    \end{bmatrix}
     =
    \underbrace{
    \begin{bmatrix}
         x_\text{e}\\
         x_\text{o}\\
    \end{bmatrix}}_{f(x)}
    +
    \underbrace{
    \begin{bmatrix}
         \cos{(\psi + \beta)}\\
         1\\
    \end{bmatrix}}_{g(x)}
    \underbrace{
    \begin{bmatrix}
        v_\text{e} & v_\text{o}
    \end{bmatrix}}_{u_t}
\end{align}
where $x_{\text{e}}$ and $v_{\text{e}}$ are the longitudinal position and velocity of the ego vehicle. Similarly, $x_{\text{o}}$ and $v_{\text{o}}$ are the longitudinal position and velocity of an observed vehicle. The velocity of an observed vehicle is considered in the control system to account for its state changes when it makes the worst-case control decision. For simplicity, the time step subscript $t$ is omitted in the above equation. 

Since the control input, $v_\text{e}$, can be possibly unsafe, our aim is to evaluate a minimum correction value $v^{\text{cbf}}$ to comply with a CBF safety constraint. Therefore, $v^{\text{safe}}$ can be defined as

\begin{equation}
    \label{eq:safe_v}
    v^{\text{safe}} = 
    \underbrace{
    \begin{bmatrix}
        v_{\text{e}} &
        v_{\text{o}}
    \end{bmatrix}}_{v}
    + 
    \underbrace{
    \begin{bmatrix}
        v^{\text{cbf}}_{\text{e}} &
        0
    \end{bmatrix}}_{v^{\text{cbf}}}
\end{equation}
In the $v_{\text{cbf}}$, the safe correction value for the observed vehicle is set to 0 as its speed cannot be controlled by the ego vehicle.

The QP optimisation problem (\ref{eq:opt_qp}) can be updated using Eq.~(\ref{eq:safe_v}) to evaluate a minimum value for $v^{\text{cbf}}$ as shown below,   
\begin{align}
\label{eq:lon_cbf_opt_qp}
\begin{split}
    v^{\text{cbf}} = & \argmin_{v^{\text{cbf}}} ||v^{\text{cbf}}||_2 + K_{\epsilon}\epsilon\\
          \text{s.t}~~& Av^{\text{cbf}} \leq b, \\
          \text{where } b =&~ [b_1, b_2,..., b_k], \\
          \text{~with}~ b_i =&~ p_i^{\text{T}}f(s_t)+ p_i^{\text{T}}(\eta - 1) s_t ~+\\ &\eta q_i + p_i^{\text{T}}g(s_t)v
\end{split}
\end{align}

Note that the above optimisation problem (\ref{eq:lon_cbf_opt_qp}) depends on a CBF constraint $h$ of the form (\ref{eq:affine_fn}). A safety constraint for an ego vehicle to maintain a safe time headway from the leading vehicle $C_{\text{ol}}$ can be defined as follows
\begin{equation}
    \label{eq:h_lon}
    h_{\text{ol}}(x_t) = \Delta x_{\text{ol}} - x^{\text{safe}}
\end{equation}
where $\Delta x_{\text{ol}}$ is the longitudinal distance from the rear end of the preceding vehicle and the front of the ego vehicle as shown in Fig.~\ref{fig:topology_a}. The $x_{\text{safe}}$ is the safe longitudinal distance that must be maintained. For an ego vehicle travelling with a velocity $v_{\text{e}}$, the safe distance threshold is calculated based on the time headway $\tau$ is defined as
\begin{equation}
\label{eq:safe_x}
    x^{\text{safe}} = \tau * v_{\text{e}}
\end{equation}

In the above Eq~(\ref{eq:safe_x}), it is important to consider that the safe distance threshold increases as the vehicle speeds up. Consequently, a vehicle may reach an unsafe state even after actuating safe control inputs, as the safe distance threshold at the current step may be higher than in the previous step. To overcome this problem, an additional buffer distance $x^{\text{buff}}$ is added to the safe distance threshold (\ref{eq:safe_x}) to account for the dynamic change, as shown below
\begin{align}
    \label{eq:x_buff}
    \begin{split}
        x^{\text{safe}} &= \tau * v_{\text{e}} + x^{\text{buff}}\\
        \text{where,~} x^{\text{buff}} &= (a^{\text{max}} + 0.1)*\Delta t~*~\tau
    \end{split}
\end{align}
where $\Delta t$ is a unit time step and $a^{\text{max}}$ is the maximum acceleration of the vehicle. A small constant $0.1$ is added to the maximum acceleration $a^{\text{max}}$ to restrict a vehicle from moving close to violating the threshold headway distance ($h_{\text{ol}}<0.001~\mathrm{m}$).

\subsubsection{Lateral control}

Safe longitudinal distance must be maintained from the adjacent vehicles in the target lane, $C_{\text{otl}}$ and $C_{\text{otr}}$, to perform a safe lane change manoeuvre and continue to maintain safety after completing the manoeuvre. Therefore, lane change rules are defined based on the safety invariance condition~(\ref{eq:cbf}) to define lateral constraints as follows
\begin{align}
    \label{eq:lateral_safe_condition}
    \begin{split}
        h_{\text{otl}}\left(x_{t+1}\right) + (\eta - 1)h_{\text{otl}}(x_t) \geq 0 \\
        h_{\text{otr}}\left(x_{t+1}\right) + (\eta - 1)h_{\text{otr}}(x_t) \geq 0
    \end{split}
\end{align}
where the CBF constraints $h_*$ are defined similarly to $h_{\text{lon}}$ from Eq.~(\ref{eq:h_lon}) by replacing the longitudinal distance $\Delta x_{\text{ol}}$ with the longitudinal distances, $\Delta x_{\text{otl}}$ and $\Delta x_{\text{otr}}$ , from vehicles $C_{\text{otl}}$ and $C_{\text{otr}}$ respectively as follows  
\begin{align}
    \label{eq:h_adjl}
    \begin{split}
        h_{\text{otl}} &= \Delta x_{\text{otl}} - x^{\text{safe}} \\
        h_{\text{otr}} &= \Delta x_{\text{otr}} - x^{\text{safe}}
    \end{split}
\end{align}
The $x^{\text{safe}}$ in the $h_{\text{otr}}$ is calculated based on the velocity of the rear adjacent vehicle $v_{\text{otr}}$ instead of the ego vehicle because the ego vehicle must ensure that a safe headway distance is maintained with respect to a rear adjacent vehicle before initiating a lane change. These constraints are applied to check safe conditions to initiate a lane change, as explained later in Section~\ref{subsec:marl_hss}.

\subsubsection{Control limits}
Given that the vehicle control inputs are subject to physical limitations, they must be constrained to avoid generating control inputs that are impossible to actuate. Let's consider the velocity range to be [$v^{\text{min}}$,~$v^{\text{max}}$].
This velocity range is integrated as a constraint in the QP-optimisation problem~(\ref{eq:lon_cbf_opt_qp}) using the following condition  
\begin{equation}
    \label{eq:control_limit_constraint}
    v^{\text{min}} \leq v^\text{cbf} + v \leq v^{\text{max}}
\end{equation}
to ensure that the safe control input is within the physical limits of the vehicle.

\subsection{MARL-HSS}
\label{subsec:marl_hss}
 
Our MARL lane change controller uses the state of the ego vehicle and the states of observed vehicles as input to evaluate a discrete lane change decision at the behavioural layer.
To execute the discrete lane change decision, a feedback controller is used at the motion planning layer to generate low-level control inputs such as acceleration $a^{\text{ll}}$ and steering angle $\psi^{\text{ll}}$. The HSS, however, uses velocity $v_*$ as control input in Eq.~(\ref{eq:nonlin_control_loncav_matrix}). Therefore, the longitudinal control input for HSS, velocity $v^{\text{ll}}_{\text{e}}$, is derived from the acceleration $a^{\text{ll}}$ based on the vehicle model (\ref{eq:kinematic_model}) as follows
\begin{equation}
    \label{eq:sim_control}
    v^{\text{ll}}_{\text{e}} =  v_\text{e} + a^{\text{ll}} \Delta t
\end{equation}
Similarly, the target velocity of the observed vehicle $v^{\text{ll}}_{\text{o}}$ is evaluated assuming a worst-case acceleration as a control input. These velocities can be combined to define a low-level control input $v^{\text{ll}} = \begin{bmatrix} v^{\text{ll}}_{\text{e}} & v^{\text{ll}}_{\text{o}} \end{bmatrix}$. 

\begin{figure}[tbp]
    \centering
    \includegraphics[width=\linewidth]{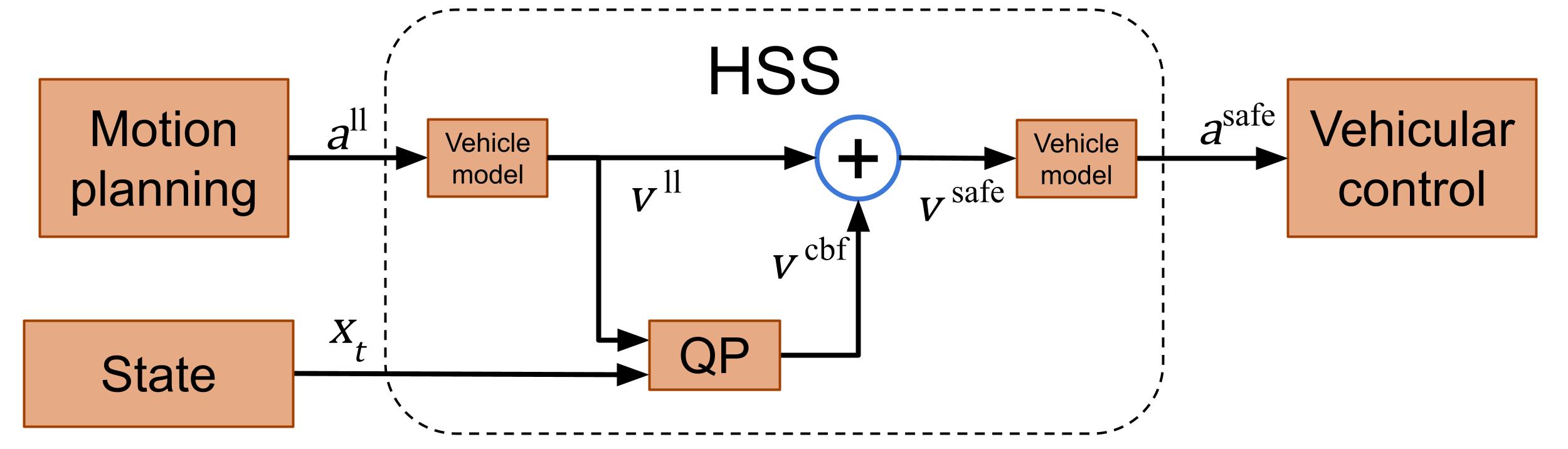}
    \caption{HSS block diagram for longitudinal control}
    \label{fig:hss_long_block}
    \vspace{-0.25cm}
\end{figure}

The longitudinal safety constraints are integrated into the feedback controller using the $v^{\text{ll}}$ in place of $v$ to solve the optimisation problem (\ref{eq:lon_cbf_opt_qp}) using QP. As this optimisation problem considers the longitudinal CBF constraint~(\ref{eq:h_lon}), the solution evaluated from QP $v^{\text{cbf}}$ is added to $v^{\text{ll}}$, as defined in Eq.~(\ref{eq:safe_v}), to evaluate safe velocity $v^{\text{safe}}$. From a component in $v^{\text{safe}}$ corresponding to the ego vehicle $v^{\text{safe}}_{\text{e}}$, the safe control input $a^{\text{safe}}$ is evaluated using the vehicle model as follows
\begin{equation}
    \label{eq:a_safe_eval}
    a^{\text{safe}} = \frac{v^{\text{safe}}_{\text{e}} - v_{\text{e}}}{\Delta t}
\end{equation}
The block diagram in Fig.~\ref{fig:hss_long_block} summarises the integration of HSS and the motion planning layer to evaluate safe longitudinal control $a^{\text{safe}}$ by overriding the possibly unsafe control input $a^{\text{ll}}$ from the motion planning layer.  

The lateral safety constraints defined in the Eq.~(\ref{eq:lateral_safe_condition}) use the low-level velocity $v^{\text{ll}}$ in place of $v$ to evaluate $h(x_{t+1})$.
If the lateral constraints are not satisfied when the behavioural layer decides to change lanes, then a new steering input $\psi^{\text{~safe}}$ is evaluated using a feedback controller to follow the current lane instead. The lateral safety constraints ensure that the lane change is initiated only when longitudinal safety constraints are maintained with both the leading and following vehicles in the target lane. As these constraints are checked during a lane change, a vehicle moves back towards the centre of the current lane if a lane change is considered unsafe midway during a manoeuvre. Therefore, this check avoids collisions between vehicles moving parallelly  the adjacent lanes. 
Since the lateral safety constraints ensure safety during the lane change and the longitudinal constraints ensure safety before and after the lane change, a vehicle is considered to make safe driving manoeuvres overall.

To integrate conditions to honour physical control limits, the velocity range [$v^\text{min}$,~$v^\text{max}$] is evaluated using a vehicle model based on the current velocity $v$ and the acceleration range [$a^{\text{min}}$,~$a^{\text{max}}$] of the CAVs, as follows
\begin{align}
    \begin{split}
        v^\text{min} = v + a^\text{min} \Delta t \\
        v^\text{max} = v + a^\text{max} \Delta t
    \end{split}    
\end{align}
This velocity range [$v^\text{min}$,~$v^\text{max}$] is used in the Eq.~(\ref{eq:control_limit_constraint}) to constrain the control limits of vehicle. 

\begin{figure}[tbp]
    \centering
    \includegraphics[width=\linewidth]{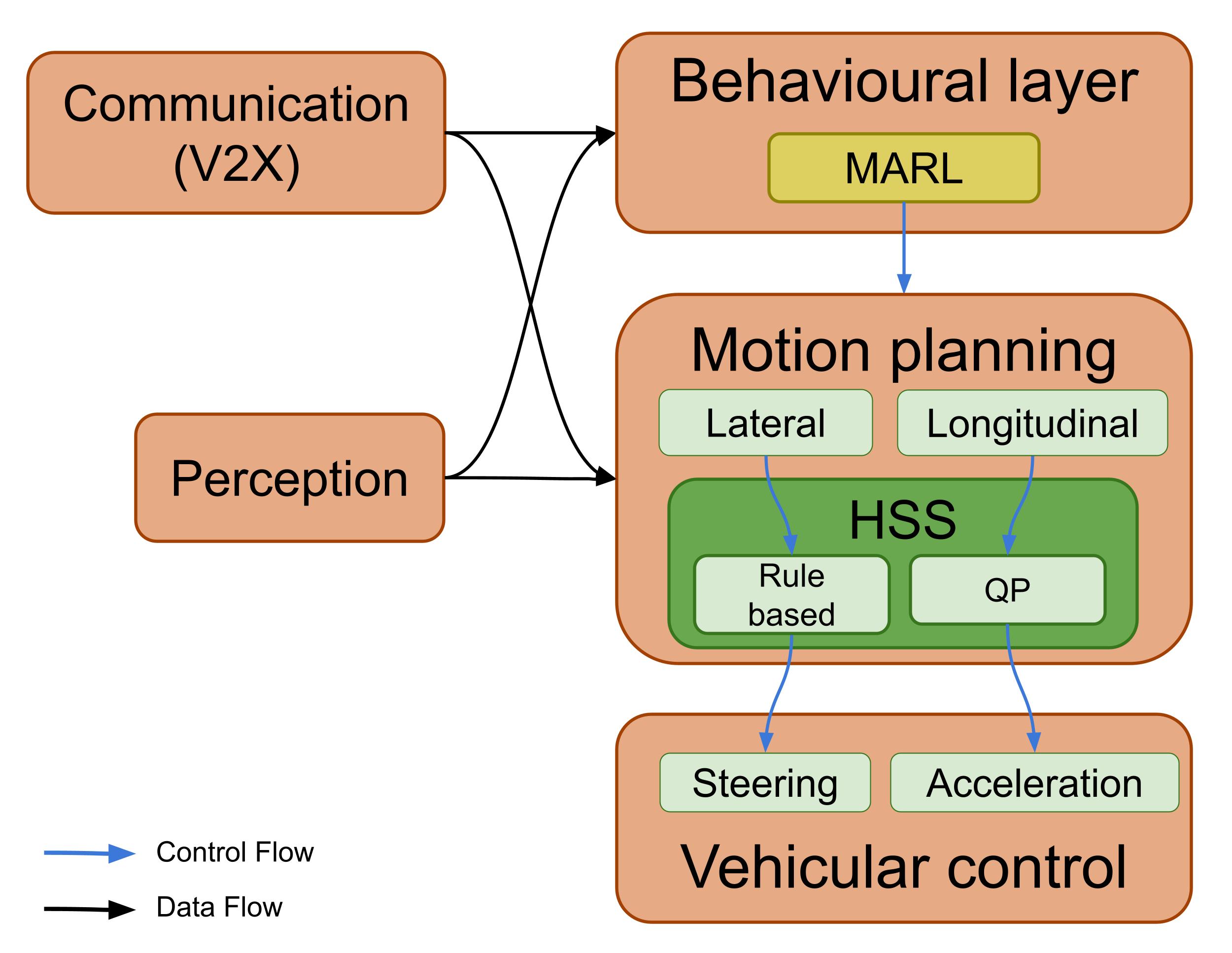}
    \caption{MARL-HSS vehicle controller architecture}
    \label{fig:marl_hss}
    \vspace{-0.25cm}
\end{figure}

To sum up, an architecture to integrate MARL (Section~\ref{subsec:marl}) and HSS (Section~\ref{subsec:hss}) is presented in Fig.~\ref{fig:marl_hss}. The longitudinal constraints in HSS ensure safe manoeuvres while following a lane before and after the lane change. During a lane change manoeuvre, lateral constraints in HSS ensure safety. Moreover, the control inputs for executing these manoeuvres are constrained to be within the physical limits of CAVs. Since all the CAVs individually evaluate safe control inputs assuming worst-case action from other vehicles, a multi-agent system of CAVs can be considered to be safe \cite{han_multi-agent_2023}.

\section{Evaluations and Results}
In this section, the proposed method, MARL-HSS, is compared with unsafe MARL defined in MARL-CAV \cite{chen_deep_2023}, which is used as the baseline. As mentioned earlier, MARL-CAV is a state-of-the-art lane change controller for making efficient lane decisions to improve traffic efficiency. 
To evaluate these methods, first a simulation setup is defined for training MARL. Then, the safety guarantee from HSS is demonstrated through minimum time headway observed while training. Next, the learning curves are illustrated to evaluate the training efficiency of MARL-HSS. Finally, MARL policies trained with MARL-HSS and the baseline are compared to analyse the trade-off between safety and efficiency.

\subsection{Evaluation scenario and simulation setup}
\begin{figure}[tbp]
\centering
    \includegraphics[width=\linewidth]{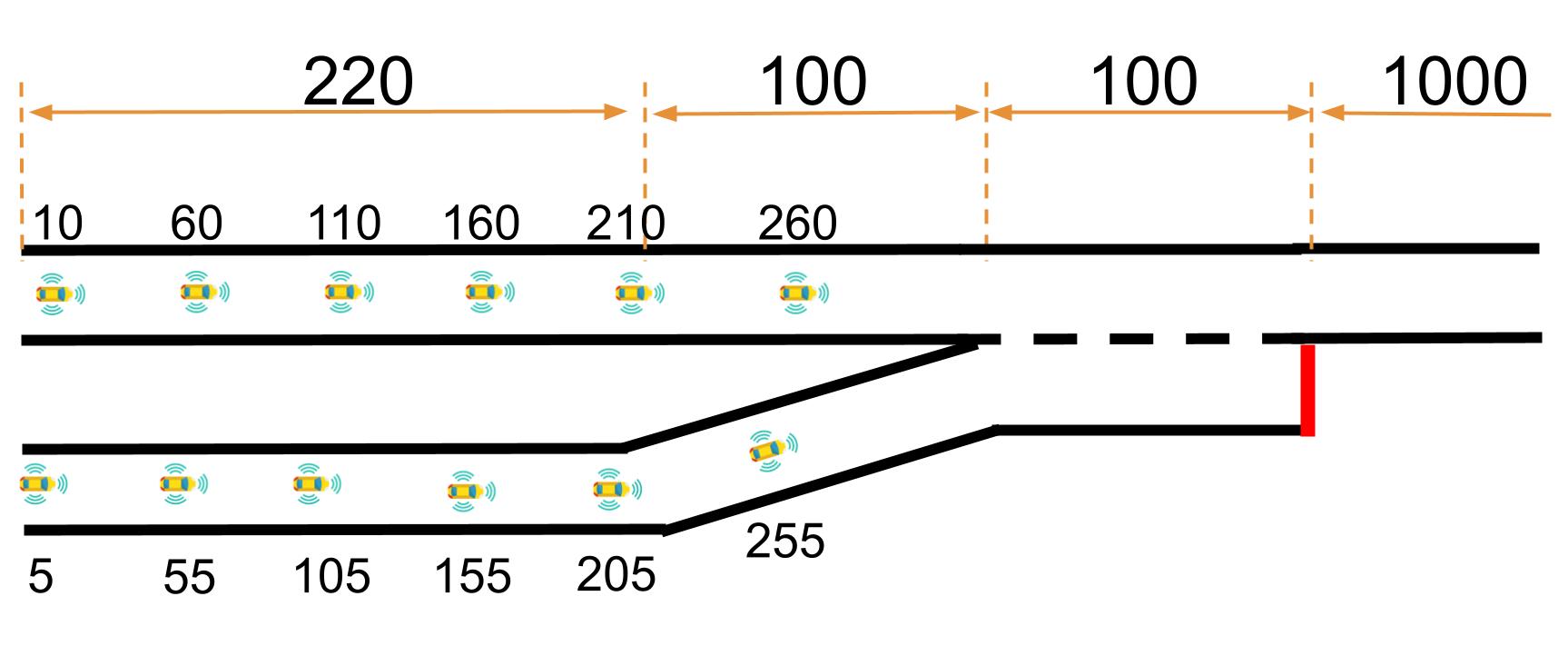}
    \caption{On-ramp merging scenario}
    \label{fig:merging_scenario}
    \vspace{-0.25cm}
\end{figure}

An on-ramp merging scenario is used to train and evaluate the MARL lane change controllers. This scenario is closely similar to the scenario used in the baseline MARL-CAV \cite{chen_deep_2023} with minor changes in the road layout, initial positions, and traffic densities. A road layout of $1420~\mathrm{m}$ is considered in our scenario consisting of $220~\mathrm{m}$ straight road to enter the merging area from the highway and a merging road followed by a $100~\mathrm{m}$ ramp that leads to a merging section of $100~\mathrm{m}$. After the merging section, the highway road is further extended by $1000~\mathrm{m}$ to allow vehicles to continue moving for 100 steps to complete a training episode. Within the first $320 ~\mathrm{m}$, CAVs are spawned at random initial positions separated by $50~\mathrm{m}$ to ensure that CAVs are in a safe state at the start, as shown in Fig.~\ref{fig:merging_scenario}.
In this merging scenario, two levels of traffic densities are considered, namely the \emph{Light traffic} consisting of 2-6 CAVs, and the \emph{Moderate traffic} consisting of 4-8 CAVs. The dense traffic with more than 8 CAVs is not considered as it may not create enough opportunities for safe lane changes.

A customised simulator is used to simulate the merging scenario by extending the \emph{highway-env} simulator \cite{leurent_environment_2018}, which is widely used to evaluate AI-based AV controllers. The custom simulator is tailored to integrate the simulation environment with the MARL algorithm (MAPPO) and HSS. 
The hyperparameters from the baseline \cite{chen_deep_2023}, such as learning rate, batch size, and reward weights ($w_*$ from (\ref{eq:reward})), are used to train MARL-HSS for a fair comparison. The policies are trained with 3 random seeds to capture the variability of rewards. During the training for 20,000 episodes, intermediate policies are evaluated using 20 episodes for every 200 training episodes. 
In the HSS, CBF constraints are configured to maintain a $0.5~\mathrm{s}$ time headway with a leading vehicle, as CAVs are expected to maintain a smaller headway compared to AVs or human-driven vehicles \cite{garg_can_2021}.
To define the strictness of CBF constraints, the coefficient of zero order control, used in the Eq.~(\ref{eq:cbf_invariance}), is configured as $\eta = 0.0325$. This value is empirically chosen using binary search to ensure safety compliance and smooth transitions in safe control inputs evaluated from HSS. 
In this simulation, the behavioural layers make control decisions at a $5~\mathrm{Hz}$ frequency, and the motion planning layers operate at a $15~\mathrm{Hz}$ frequency. Therefore, MARL must evaluate the lane change decision within $0.2~\mathrm{s}$. Moreover, the motion planning layer consisting of the feedback controller and HSS must evaluate the safe control action within $0.0667~\mathrm{s}$.
Both the baseline and MARL-HSS lane change controllers were trained in a shared Ubuntu 24.04 server with AMD EPYC 9654 processor, 4$\times$40 GB GPUs (Nvidia-L40), and 1.5 TB memory. They only used, however, approximately  1.5 GB GPU memory and 3 GB RAM memory.

\subsection{Safety evaluation}

\begin{figure}[tbp]
\centering
    \includegraphics[width=\linewidth]{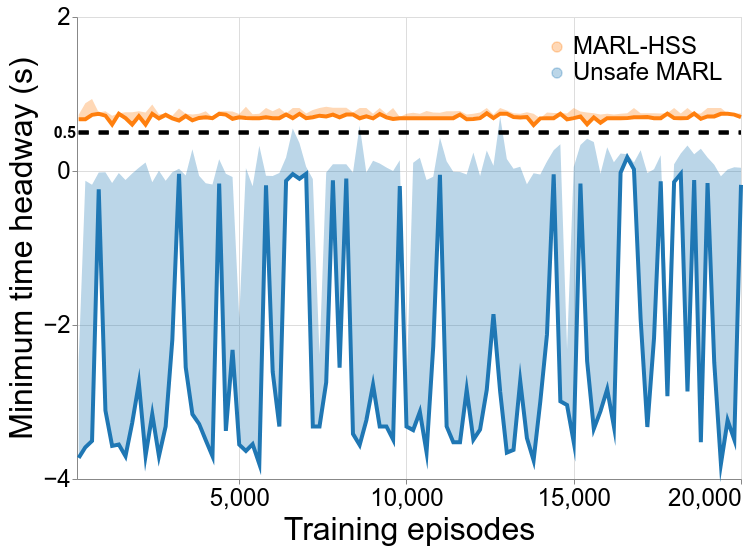}
    \caption{Minimum time headway observed during the MARL training in moderate traffic. The shaded region illustrates the range of minimum time headway observed over 3 random seeds.}
    \label{fig:min_hewadway_td2}
    \vspace{-0.25cm}
\end{figure}

In this subsection, the safety guarantee from HSS is analysed in moderately dense traffic that offers challenging lane change scenarios with limited opportunities for safe lane changes. In such scenarios, MARL policy will push the CAVs towards violating the time headway safety constraint ($0.5~\mathrm{s}$) to achieve higher rewards. 
Typically, time headway is a positive number indicating a time gap between two vehicles. However,  when a collision occurs, negative values are observed in the simulation. The minimum time headway is captured by checking the time headway of each CAV at each step of the episodes used for the intermediate evaluations while training. The minimum time headways observed with unsafe MARL and MARL-HSS are compared in Fig.~\ref{fig:min_hewadway_td2}. 
With the baseline controllers, CAVs frequently violate the safety constraint, leading to crashes. These violations can be avoided with HSS integration as MARL-HSS does not cross the safety constraint of $0.5~\mathrm{s}$ time headway.
The proposed safety shield, therefore, ensures that the dynamic safety constraint defined using time headway is enforced strictly. 

\subsection{MARL training}
\begin{figure}[tbp]
\centering
    \begin{subfigure}{\linewidth}
    \begin{subfigure}[t]{0.48\linewidth}
        \includegraphics[width=\linewidth]{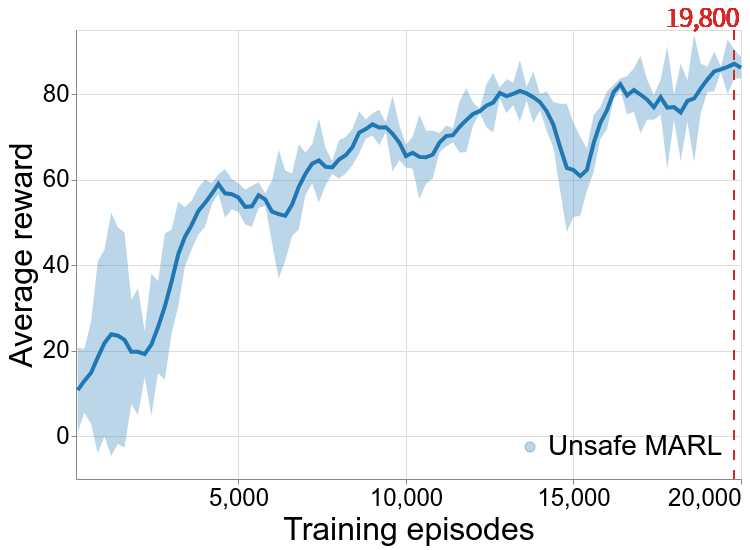}
    \end{subfigure}
    \begin{subfigure}{0.48\linewidth}
        \includegraphics[width=\linewidth]{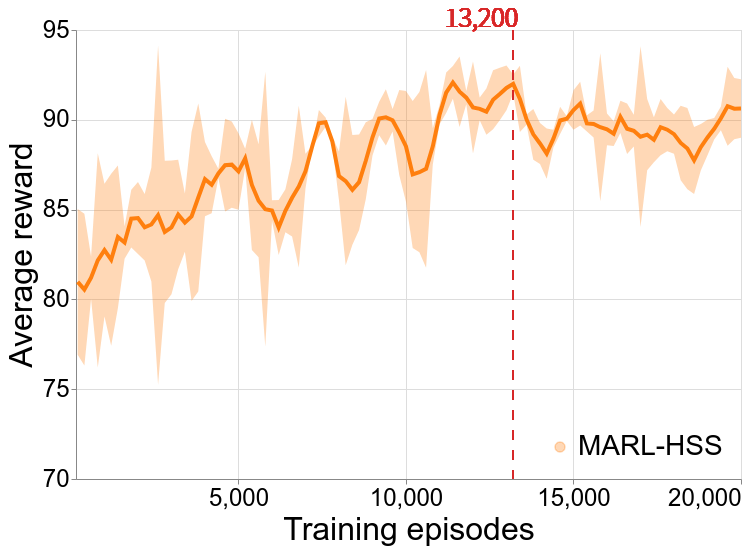}
    \end{subfigure}
    \caption{Light traffic}
    \label{fig:reward_curve_td1}
    \end{subfigure}
    \begin{subfigure}{\linewidth}
    \begin{subfigure}[t]{0.48\linewidth}
        \includegraphics[width=\linewidth]{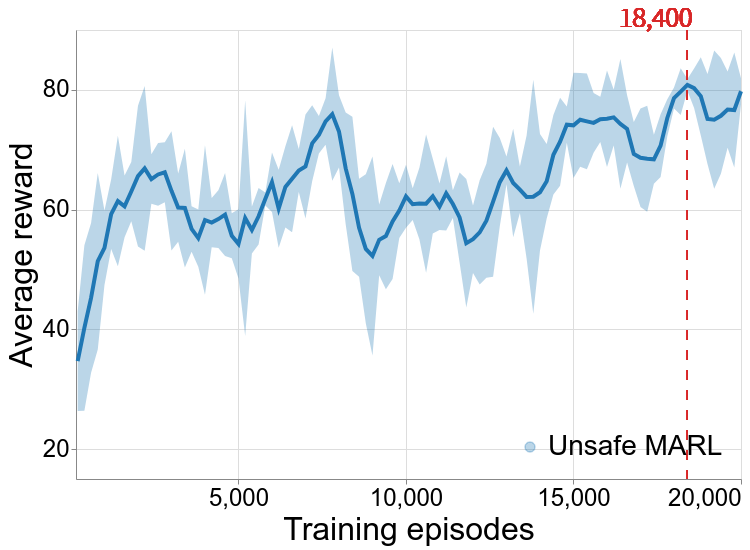}
    \end{subfigure}
    \begin{subfigure}{0.48\linewidth}
        \includegraphics[width=\linewidth]{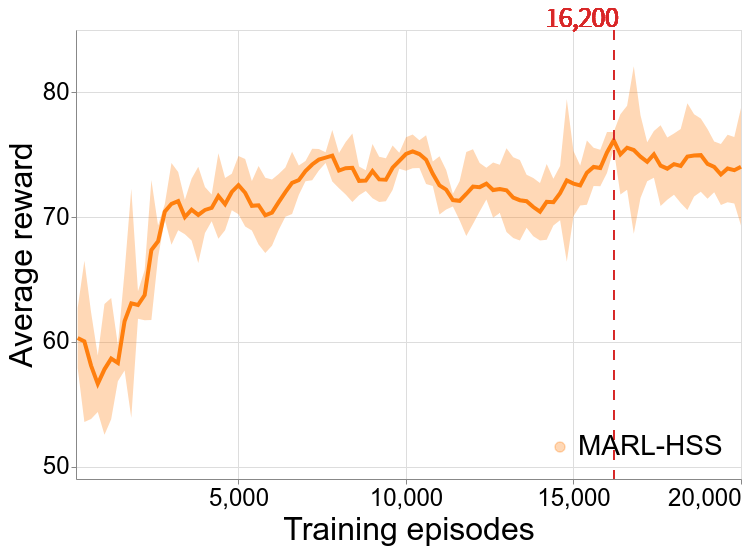}
    \end{subfigure}
    \caption{Moderate traffic}
    \label{fig:reward_curve_td2}
    \end{subfigure}
    \caption{Comparison of the reward curves during trainings in \emph{Light} and \emph{Moderate} traffic levels. The spread indicates the standard error observed over 3 random seeds. The curves are smoothed using an average over a window of 5 evaluation epochs}
    \label{fig:reward_curves}
\end{figure}

The reward curves from MARL-HSS are compared with the baseline in Fig.~\ref{fig:reward_curves} to evaluate the training performance. The reward curves illustrate a higher sample efficiency of the proposed method compared to the baseline. As light traffic provides sufficient opportunities for safe lane changes, our approach comfortably achieves higher rewards after 13,200 episodes of training, as shown in Fig.~\ref{fig:reward_curve_td1}. Without a safety shield, the unsafe MARL needs to explore more episodes compared to MARL-HSS to achieve higher rewards. Therefore, it achieves higher rewards towards the later stages of training after 19,800 episodes. Similarly, while the proposed method learns the best policy after 16,200 episodes, the baseline requires 18,400 episodes of training, as illustrated in Fig.~\ref{fig:reward_curve_td2}.

Importantly, the proposed method learns a policy that is more stable than the baseline, as shown by unsafe MARL policies having higher standard errors than MARL-HSS for both traffic levels. Moreover, our method achieves more consistent rewards throughout the training process in both traffic levels compared to the unsafe MARL's rewards that drop noticeably.

\subsection{MARL policy evaluation}
In this subsection, the best-performing policies are evaluated using 100 \emph{policy evaluation episodes}, which are different from the episodes used for intermediate evaluation while training. The policies are compared on the basis of speed ($\mathrm{m/s}$) to evaluate traffic efficiency, and crash counts are considered for safety evaluation. The average values of these parameters over the policy evaluation episodes are presented in Table~\ref{tab:policy_evaluation}.

\begin{table}[tbp]
\caption{Policy evaluation of MARL-HSS and unsafe MARL in different traffic levels using 100 episodes}
\vspace{-0.1cm}
\begin{center}
\begin{tabular}{llccccccc}
\toprule
\textbf{} & \begin{tabular}[c]{@{}c@{}}Traffic \\ level\end{tabular} & \begin{tabular}[c]{@{}c@{}}Training \\ episodes \end{tabular} & \begin{tabular}[c]{@{}c@{}} Speed \\ ($\mathrm{m/s}$) \end{tabular} & \begin{tabular}[c]{@{}c@{}} Crash \\ count\end{tabular} \\ \midrule
\textbf{MARL-HSS} & Light & 13,200 & \textbf{28.36} & \textbf{0}   \\
& Moderate & 16,200 & 26.54 & \textbf{0}  \\
\textbf{Unsafe MARL} & Light & 19,800 & 27.70 & 6.3 \\ 
& Moderate  & 18,400 & \textbf{27.32} & 5 \\ 
\bottomrule
\end{tabular}%
\vspace{-0.75cm}
\label{tab:policy_evaluation}
\end{center}
\end{table}

The results in Table~\ref{tab:policy_evaluation} show that MARL-HSS balances a trade-off between safety and efficiency. With the MARL-HSS policy, zero crashes are observed in both traffic levels, which reaffirms safety guarantees. However, mean crash counts of $6.3$ and $5$ are observed with unsafe MARL in light and moderate traffic, respectively. In a light traffic density, the policy with the proposed safety shield marginally improves the traffic efficiency with a $2.38\%$ increase in the average speed. A comparable traffic efficiency is observed between the proposed method and the baseline in moderate traffic with a marginal difference of $0.78~\mathrm{m/s}$ in the average speeds. With closely similar average speeds, MARL-HSS is considered to achieve traffic efficiency of the baseline while respecting safety constraints.

\section{Conclusion}

In this article, we propose a decentralised safety shield, HSS, to ensure safety based on dynamic constraints. This safety shield formulates CBF constraints for longitudinal and lateral control to override possibly unsafe control inputs with a minimum correction to ensure safe manoeuvres.    
To integrate HSS into the MARL lane change controller, we define the MARL-HSS architecture. In this architecture, the behavioural control decisions from the lane change controller are translated into safe control inputs at the motion planning layer.
Experimental evaluation shows that the dynamic safety constraints are strictly enforced, preventing unsafe actions even when MARL prioritises higher rewards. 
Moreover, the proposed lane change controller learns a stable policy with less variability in the training. 
While the policy trained with MARL-HSS outperforms the state-of-the-art baseline in light traffic density, it achieves comparable traffic efficiency in moderate traffic. 
Notably, MARL-HSS balances a trade-off between safety and traffic efficiency by ensuring safety across different traffic densities without significantly compromising on the average speed.

MARL-HSS could be extended in the future to support mixed traffic, by considering uncertainties associated with vehicles of variable levels of autonomy and connectivity. Moreover, exploiting CAV's ability to collaboratively decide control actions in the safety shield could potentially further improve traffic efficiency while ensuring safety.

\section*{Acknowledgment}
The authors wish to thank the editors and anonymous reviewers for their helpful suggestions.
This publication has emanated from research conducted with the financial support of Taighde Éireann – Research Ireland under Grant numbers 18/CRT/6222 at Centre for Research Training in Advanced Networks for Sustainable Societies (ADVANCE CRT) and 13/RC/2077\_P2 at CONNECT: the Research Ireland Centre for Future Networks. 

\bibliographystyle{IEEEtran}

\end{document}